\documentclass[%
 aip,
 amsmath,amssymb,
 preprint,%
]{revtex4-1}

\usepackage{graphicx}
\usepackage{xcolor}
\usepackage{dcolumn}
\usepackage{bm}

\usepackage[utf8]{inputenc}
\usepackage[T1]{fontenc}
\usepackage{mathptmx}
\usepackage{SIunits}
\usepackage[normalem]{ulem}

\newcommand{\wse}{WSe$_2$~}

\begin{document}

\preprint{AIP/123-QED}

\title{Spectral and spatial isolation of single \wse quantum emitters using hexagonal boron nitride wrinkles}

\author{Rapha\"el S. Daveau}
 \affiliation{School of Applied and Engineering Physics, Cornell University, Ithaca, New York 14853, USA}
\author{Tom Vandekerckhove}%
 \affiliation{School of Applied and Engineering Physics, Cornell University, Ithaca, New York 14853, USA}
\author{Arunabh Mukherjee}
 \affiliation{The Institute of Optics, University of Rochester, Rochester, New York 14627, USA}
  \author{Zefang Wang}
 \affiliation{School of Applied and Engineering Physics, Cornell University, Ithaca, New York 14853, USA}
  \author{Jie Shan}
 \affiliation{School of Applied and Engineering Physics, Cornell University, Ithaca, New York 14853, USA}
 \affiliation{Laboratory of Atomic and Solid State Physics, Cornell University, Ithaca, New York 14853, USA}
  \affiliation{Kavli Institute at Cornell for Nanoscale Science, Ithaca, New York 14853, USA}
  \author{Kin Fai Mak}
 \affiliation{School of Applied and Engineering Physics, Cornell University, Ithaca, New York 14853, USA}
 \affiliation{Laboratory of Atomic and Solid State Physics, Cornell University, Ithaca, New York 14853, USA}
  \affiliation{Kavli Institute at Cornell for Nanoscale Science, Ithaca, New York 14853, USA}
 \author{A. Nick Vamivakas}
 \affiliation{The Institute of Optics, University of Rochester, Rochester, New York 14627, USA}
\author{Gregory D. Fuchs}%
 \email{gdf9@cornell.edu}
\affiliation{School of Applied and Engineering Physics, Cornell University, Ithaca, New York 14853, USA}
\affiliation{Kavli Institute at Cornell for Nanoscale Science, Ithaca, New York 14853, USA}

\date{\today}

\begin{abstract}
Monolayer WSe$_2$ hosts bright single-photon emitters. Because of its compliance, monolayer \wse conforms to patterned substrates without breaking, thus creating the potential for large local strain, which is one activation mechanism of its intrinsic quantum emitters. Here, we report an approach to creating spatially and spectrally isolated quantum emitters from \wse monolayers with few or no detrimental sources of emission.  We show that a bilayer of hexagonal boron nitride (hBN) and \wse placed on a nanostructured substrate can be used to create and shape wrinkles that communicate local strain to the \wse, thus creating quantum emitters that are isolated from substrate features. We compare quantum emitters created directly on top of substrate features with quantum emitters forming along wrinkles and find that the spectra of the latter consist of mainly a single peak and a low background fluorescence. We also discuss possible approaches to controlling emitter position along hBN wrinkles.
\end{abstract}

\maketitle

Solid-state quantum emitters (QEs) such as semiconductor quantum dots and color centers in solids have promising applications in quantum information science, optoelectronics and nano-sensing  because of their versatility, stability and sensitivity \cite{obrien2009photonic, gisin2002quantum, kimble2008quantum, aharonovich2016solid}. Monolayer transition metal dichalcogenides (TMDs) are direct bandgap semiconductors with strong light-matter interactions, large excitonic effects and a valley degree of freedom that locks excitons to a given photon helicity \cite{mak2016photonics}. Quantum emitters in TMDs that inherit valley properties could be used for the generation of single photons with orthogonal polarization for polarization-encoded flying qubits \cite{schaibley2016valleytronics}. One TMD, tungsten diselenide (WSe$_2$), hosts bright and stable QEs \cite{srivastava2015optically, he2015single, koperski2015single, chakraborty2015voltage}, which have been used in a broad range of experiments that shows their potential: fine-structure splitting manipulation with electric field \cite{chakraborty2019electrical}, linear-to-circular helicity conversion with magnetic field \cite{he2015single}, quantum confined Stark tuning of excitons \cite{chakraborty2017quantum}, and single charge injection to trion state \cite{chakraborty20183d,brotons2019coulomb}. 

While most reported emitters in \wse monolayers are located at random, a few reports demonstrate quasi-deterministic activation of QEs via local strain by placing the \wse  monolayer over a substrate patterned with nanopillars \cite{branny2017deterministic, palacios2017large}, a slotted waveguide \cite{kern2016nanoscale} or hexagonal boron nitride nano-bubbles \cite{shepard2017nanobubble}. Based on these works, it has become clear that strain plays a role in activating \wse QEs, although their precise nature -- whether excitonic or defect-bound \cite{branny2017deterministic, zheng2019point} -- is still being actively investigated. There is still no reliable method to strain-engineer single QEs in WSe$_2$ deterministically. In most reports, the strain in \wse monolayers that is created at random or at a pillar apex is complex and non-uniform, resulting in the creation of multiple QEs per site. Eventually this results in the excitation and emission from several QEs at once leading to degraded single photon purity and the need for tunable spectral filtering of specific emitters. 

Here, we report on an alternative approach to create spatially isolated single QEs in \wse  monolayers with the help of hexagonal boron nitride (hBN). Unlike previous \wse QE studies involving hBN as an insulating layer in heterostructures including graphene \cite{chakraborty20183d, chakraborty2017quantum, chakraborty2019electrical}, we use hBN as a strain buffer from a nanostructured substrate to control the amount of strain applied to WSe$_2$. By stacking \wse with sub-10-nm hBN, wrinkles nucleate from a nanostructured substrate and create local strain in \wse at a spatial location away from the substrate pattern. We find that these wrinkle points frequently host a \emph{single} QE with very low background. These two properties -- number of QEs created per site and the peak-to-background ratio at the emitter frequency -- are important figures-of-merit to quantify since spectral overlap from other emitters and background light on top of a QE of interest are intrinsically detrimental to the single-photon purity of the source. We extract a peak-to-background ratio as high as $0.99\pm0.05$ for wrinkle-based QEs with an average of $0.88\pm0.03$ for a 3~nm spectral width. In comparison, pillar-based QEs have an average peak-to-background ratio of $0.74\pm0.03$ due to the typical formation of a much richer spectrum composed of several emitter lines and a larger background. We also find that using a thin hBN capping layer above \wse reduces spectral wandering and blinking. Finally, we discuss approaches to controllably introduce local wrinkles in hBN in desired locations, thus enabling deterministic positioning of single \wse QEs. In this vein, we present a sample patterned in different designs to exemplify that many strain profiles are suitable to activate QEs and that nanopillars are not the obvious way to activate \wse QEs.

\section*{Nanopillar versus wrinkle-based quantum emitters}

Strain engineering of \wse monolayers is a proven method for creating QEs on demand and with position control \cite{branny2017deterministic, palacios2017large, cai2018radiative}. Previous methods have employed \wse monolayers strained directly on top of a nano-patterned substrate to create QEs. Our approach uses a nano-patterned substrate as an indirect means to communicate strain to \wse via a thin hBN layer. By conforming to the uneven substrate, the thin hbN layer supports wrinkles and small height variations, which strains \wse differently than if \wse were to conform to the lithographically defined structures by itself. In the first part of this work, we use a substrate patterned with nanopillars following the fabrication recipe of Proscia \emph{et al.} \cite{proscia2018near} and make a comparative study of emitters forming directly on the pillars versus along hBN wrinkles that propagate between pillars. 

\begin{figure*}[t!]
  \centering
  \includegraphics[width=\linewidth]{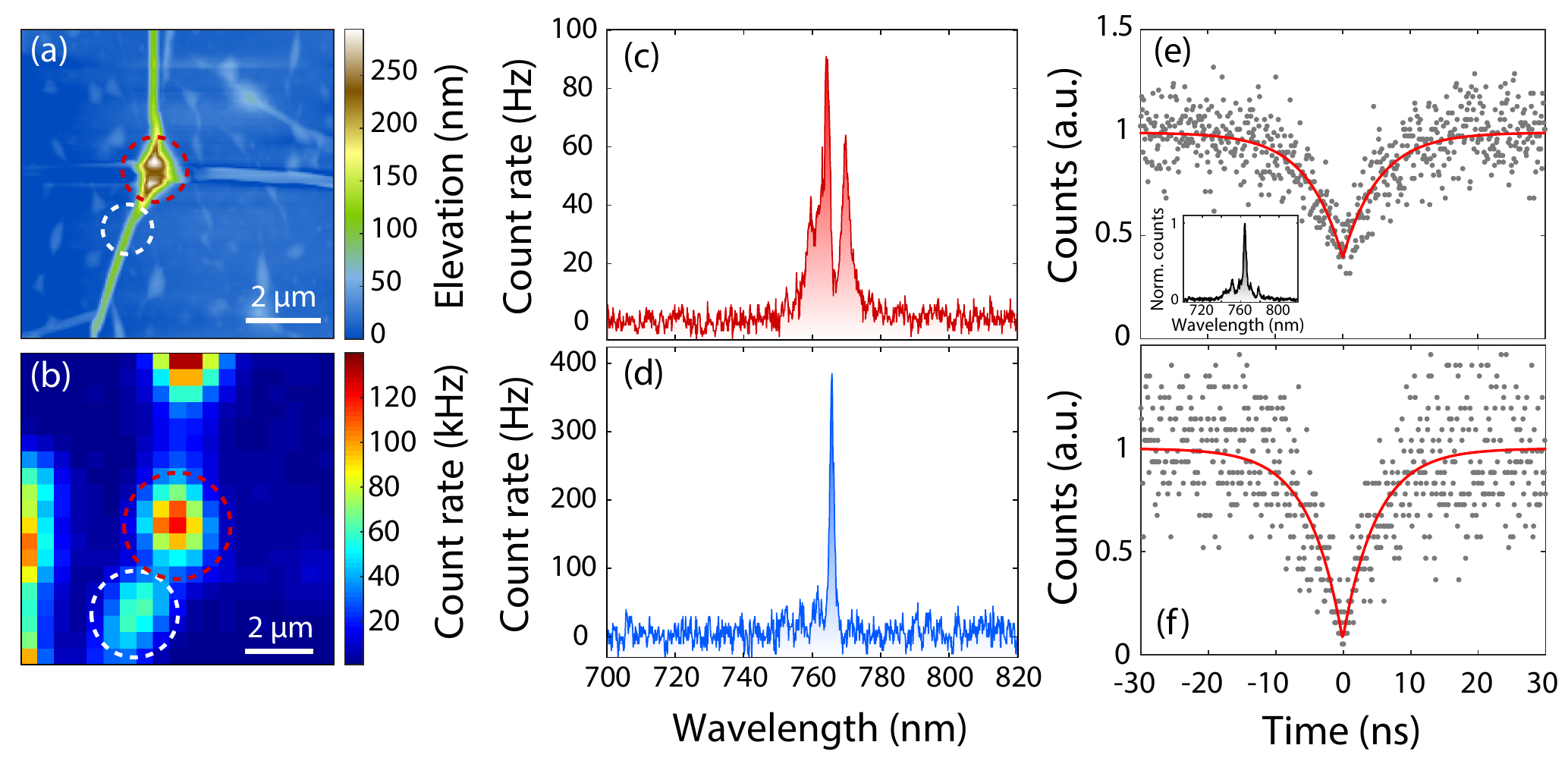}
  \caption{Comparison between pillar emission versus wrinkle emission on a hBN/\wse device (sample~1). a) Atomic force microscopy image showing a pillar and wrinkles shooting off the pillar. Wrinkles form in the thin hBN layer (thickness 9~nm). The stacked \wse conforms to the hBN. Circled are the two spots of interest: top of the pillar (red) and wrinkle kink (white). b) Photoluminescence map of the area while exciting at 637~nm with 500~nW laser power (power density is 50~W/cm$^2$). c) Spectrum collected from the center of the nanopillar, see red circle in a). d) Spectrum collected from the wrinkle as indicated by the white circle in a). Laser power is 100~nW. e) Auto-correlation histogram $g^{(2)}(\tau)$ of a QE (spectrum shown as inset) from the center of a pillar excited with $P=500$~nW. The QE line is selected through a 3-nm bandpass filter. $g^{(2)}(0) = 0.393 \pm 0.015$. f) Auto-correlation histogram of the wrinkle QE shown in d) excited with $P=500$~nW, filtered through a 3-nm bandpass filter. The extracted $g^{(2)}(0) = 0.087 \pm 0.004$. }
  \label{fig:fig1}
\end{figure*}

The nanopillars are fabricated from an SiO$_2$(500~nm)/p-Si(500~\micro m) substrate via electron-beam lithography (JEOL 6300) using M-aN 2403 negative resist \cite{proscia2018near}. The nanopillars (and other structures) are etched into SiO$_2$ and are cylindrical with a diameter of 200~nm and height of 300~nm.  The spacing between pillars is 4~\micro m, which we found is phenomenologically the minimum spacing that allows the WSe$_2$/hBN stack to both conform to the pillars and adhere to the substrate via Van der Waals forces. To place an exfoliated \wse  monolayer deterministically over the fabricated nanopillars, we employ a polycarbonate (PC) on polydimethylsiloxane (PDMS) stamp technique \cite{wang2013one}, which is commonly used to make heterostructures of 2D materials. The PC is used to pick up \wse and hBN (or vice versa) successively and  melted at 180~$^\text{o}$C on top of the target substrate. The melted PC is finally dissolved in chloroform for 10 minutes. The hBN/\wse stack is transferred such that the \wse monolayer is in contact with a Au electrode. A small bias of 1~V is applied accross the SiO$_2$ layer between the \wse monolayer and the back silicon layer, which is used to passivate the electrostatic environment and stabilize QEs. In this section, we study two samples made from the same substrate: sample~1 contains a hBN(bottom)/\wse stack and sample~2 contains a WSe$_2$/hBN(top) stack.

We study the samples using a homebuilt confocal microscope setup with a 637~nm continuous wave laser focused with a 50x Olympus ($NA=0.7$) to the diffraction limit ($<1$~$\mu$m). The sample is placed in a helium flow cryostat at 10~K. The laser light is filtered out in collection with a long-pass filter and sent onto silicon avalanche photodiodes (APDs) for time-resolved measurements or a spectrometer (Princeton Acton SP-2500) with focal length $f=500$~mm and a 300~g/mm grating. Single QE lines are filtered through a Semrock tunable filter with a 3~nm transmission bandwidth.

Previous studies using nanopillars to activate QEs show that \wse strained by a nanopillar apex become the brightest centers of photoluminescence on the monolayer \cite{kumar2015strain,palacios2017large,branny2017deterministic}. The spectra associated with these bright spots are composed of several sharp lines and a broad background of weakly bound excitons (720-780~nm). In this present work, we are able to reproduce similar results when we study the bright emission centers directly on top of nanopillars. Most remarkably, we identify that one emission center along a hBN wrinkle corresponds to a single QE with low surrounding background emission. Figure~\ref{fig:fig1}(a) plots an atomic force microscopy (AFM) image of sample~1 consisting of a hBN/\wse heterostructure on a SiO$_2$ nanopillar substrate. The photoluminescence (PL) map of this sample area obtained after exciting at 637~nm with 500~nW laser power is plotted in Fig.~\ref{fig:fig1}(b). The brightest emission is recorded on top of pillars (center and top of the image) and at the folded monolayer edge (left). A fluorescent spot is also visible along a hBN wrinkle, as indicated with a white dashed circle in Fig.~\ref{fig:fig1}(a-b). The spectrum from the center of the nanopillar is shown in Fig.~\ref{fig:fig1}(c) and features a broad emission peak from 755~nm to 775~nm, on top of which two peaks may be identified. Figure~\ref{fig:fig1}(d) shows a spectrum collected from the wrinkle shooting off the nanopillar (see white circle). This spectrum features a comparably single sharp peak at 765.8~nm with minimal background emission surrounding the peak. The AFM image reveals a kink in the wrinkle, which can explain the creation of a QE at this specific site. At this location, the wrinkle is 90~nm high and has a full-width at half-maximum (FWHM) of 160~nm. The collected light from this spot is filtered through a 3~nm bandpass filter and sent to a Hanbury-Brown-Twiss interferometer for auto-correlation measurement. The time correlation between the two APD signals for the wrinkle QE is plotted as $g^{(2)}(\tau)$ in Fig.~\ref{fig:fig1}(f). The data is fitted with $g^{(2)}(\tau) = 1 - A_1\exp(-|\tau|/t_1)$, where $t_1$ and $A_1$ is a characteristic time and amplitude of photon antibunching, respectively. From the fit we extract $g^{(2)}(0) = 0.087 \pm 0.004$, unequivocally demonstrating the single-photon nature of the emission. Figure~\ref{fig:fig1}(e) shows the $g^{(2)}(\tau)$ of a filtered QE collected from a pillar (spectrum shown as inset), from which we extract $g^{(2)}(0) = 0.393 \pm 0.015$. The detrimental effect of spectral proximity to other emitters and background emission collected with the QE clearly impacts the single-photon purity. In the following section, we quantify the amount of unwanted background emission associated with several QEs. This sample provides a first insight into the difference in spectral quality between pillar and wrinkle QEs; an effect that we study extensively in the next sample where we observe reproducible behavior. 

\begin{figure}[t!]
  \centering
  \includegraphics[width=0.5\linewidth]{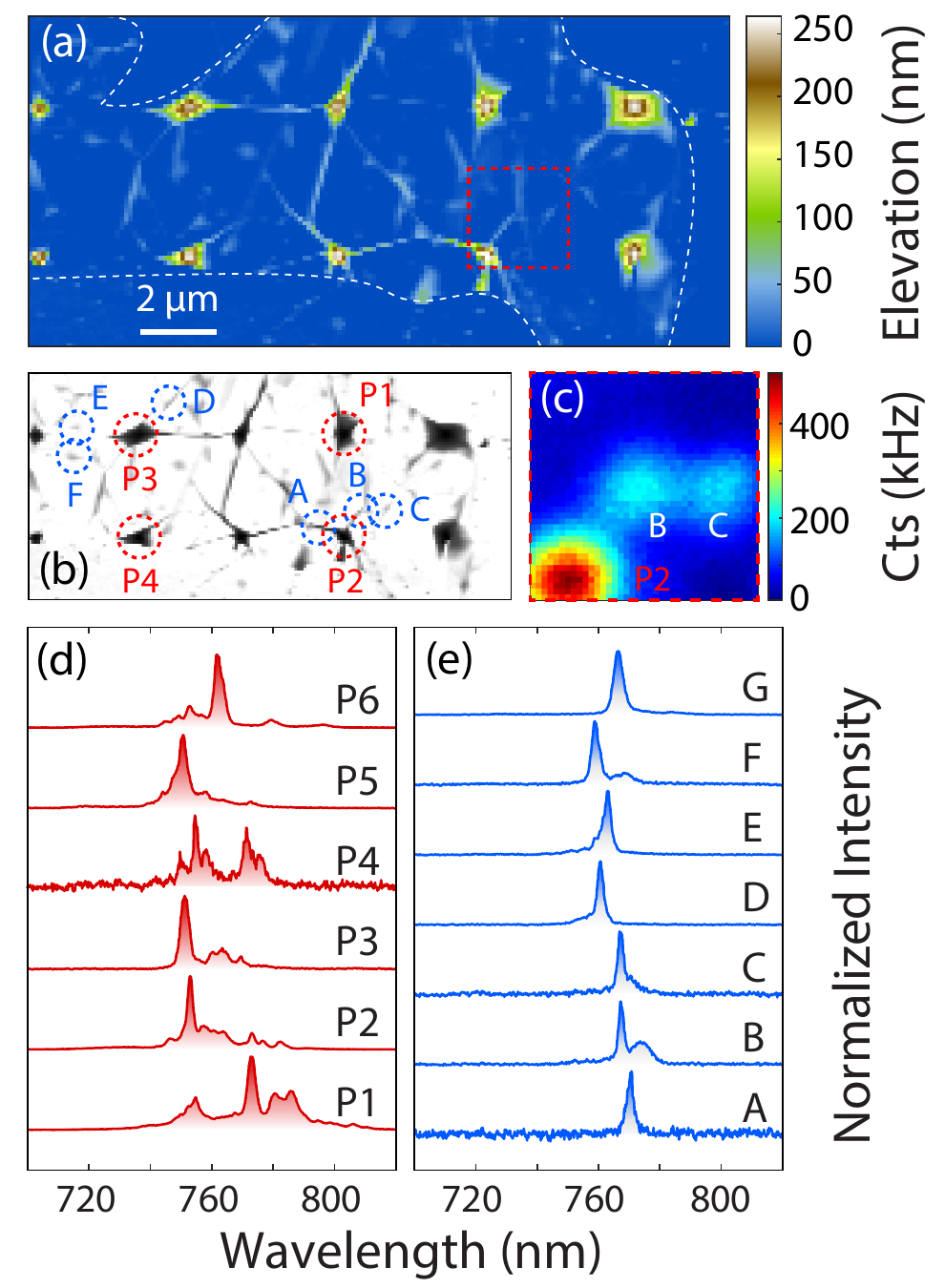}
  \caption{Comparison between pillar emission versus wrinkle emission on a WSe$_2$/hBN  device (sample~2). a) Atomic force microscopy image of the sample. The white dashed lines show the edges of the hBN flake. b) Black and white rendering of a) used as a guide to label pillars and hBN wrinkles. c) Photoluminescence map of the sample next to pillar P2. The two emission spots in the center stem from wrinkles and/or nano-bubbles. Laser wavelength is 637~nm and power is 8~\micro W. d) Set of spectra captured at the center of several nanopillars, which also correspond to the brightest PL centers under \micro -PL scans. The spectra feature one or several sharp lines and a broad background emission. e) Set of spectra collected from wrinkles and nano-bubbles between nanopillars. The spectra consistently feature a single peak and the background level is significantly less than at the nanopillars. For labels refer to b).}
  \label{fig:fig2}
\end{figure}

We now study sample~2 where the hBN layer is on top of \wse instead of the bottom. As in the previous case, single QEs are found on the pillars as well as between pillars where hBN forms wrinkles or nano-bubbles \cite{shepard2017nanobubble, darlington2020imaging}. Figure~\ref{fig:fig2}(a) shows an AFM image of the sample where 6 nanopillars are investigated. All  on-pillar and off-pillar emission centers are labeled in Fig.~\ref{fig:fig2}(b). Again, a PL-map -- recorded from the area inside the red square box in Fig.~\ref{fig:fig2}(a) -- shows that pillars emit the most light while there are other emission centers away from the pillars, see Fig.~\ref{fig:fig2}(c). We compare emission spectra of 6 on-pillar emitters and 7 off-pillar emitters in Figs.~\ref{fig:fig2}(d-e). The spectra collected from the top of the pillars show one or several lines attributed to QEs and a broad emission background. We quantify the peak-to-background ratio $r_0 = I_\text{peak}/(I_\text{peak} + I_\text{background})$ for a 3~nm spectral window, corresponding to our experimental conditions. To extract $I_\text{peak}$ and $I_\text{background}$, each spectra is fitted with the following function $a_0/((\lambda-\lambda_0)^2 + \gamma^2) + b_0(\lambda)$, where $a_0$ is proportional to the peak intensity, $\lambda_0$ is the peak center wavelength, $2\gamma$ is the peak full-width at half maximum (FWHM) and $b_0(\lambda)$ is the wavelength-dependent background level. $I_\text{peak}$ is computed as the integrated intensity of the peak and $I_\text{background}$ is the integrated $b_0(\lambda)$, both of which limited to a 3~nm spectral window around the emitter center wavelength. For a graphical representation of how the ratio is extracted from a spectrum, refer to Fig.~\ref{fig:figSI1} in Appendix~\ref{sec:SI:r0}. Quantum emitters stemming from pillars have an average $r_0$ of $0.74\pm0.03$. That is, a fourth of the filtered emission comes from sources other than the single-photon source of interest. This background emission cannot be experimentally separated from the single photons and such sources will intrinsically have a low single-photon purity and limited applicability \cite{gisin2002quantum, broome2013photonic}. In Fig.~\ref{fig:fig2}(e), we sample spectra that are collected along wrinkles that form away from the pillars. The spectra consistently display a single peak, i.e., a single QE probed at a time, while the background level is significantly reduced compared to the spectra of Fig.~\ref{fig:fig2}(b). The average $r_0$ for wrinkle-based QEs is $0.88\pm0.03$, meaning one photon in eight comes from unwanted sources in the filtered signal, which is a 2$\times$ improvement. Four emitters out of seven (labeled A, E, F and G), have $r_0 > 0.9$ with the highest value being $0.99\pm0.05$, all allowing for high single-photon purity. We speculate that the variation in $r_0$ is due to the different wrinkle sizes and morphology, which influence the confining potential and the presence of weakly bound excitons that contribute to background fluorescence. We also note that there are multiple instances in the literature that report spectrally isolated single quantum emitters, formed at random over a thick exfoliated hBN substrate \cite{kumar2016resonant} or from high aspect ratio nanopillars with sub-100~nm diameter \cite{palacios2017large}. 

The peak-to-background ratio estimation can be extended to the full spectrum as a way to probe the overall amount of background and other emitters collected in the signal along with the QE of interest. In practice, we compare a QE peak intensity to the sum of all pixels in the collected spectrum. On average, the main QE peak makes 60~\% of the signal along wrinkles and 26~\% on pillars. That is, for equal peak intensity, single photons from wrinkle-based QEs are collected with roughly 4 times as little unwanted emission in the unfiltered spectrum compared to pillar QEs. In the best case, we find that one wrinkle QE has a full-spectrum peak-to-background ratio of $0.96\pm0.03$. Further improvement of this ratio could allow for direct use of the single-photon source without spectral filtering, which is currently a default setting for semiconductor sources \cite{Lodahl2015interfacing}. A known way to clean a QE's spectrum is to use quasi-resonant excitation, whereby the QE is excited through a higher energy state and other emitters do not respond to the same excitation energy \cite{kumar2016resonant}. This requires a tunable laser close to the QE ground state energy and most often a cross-polarization excitation scheme.

\section*{Influence of Hexagonal boron nitride on the spectral stability of \wse quantum emitters}

\begin{figure}[b!]
  \centering
  \includegraphics[width=0.5\linewidth]{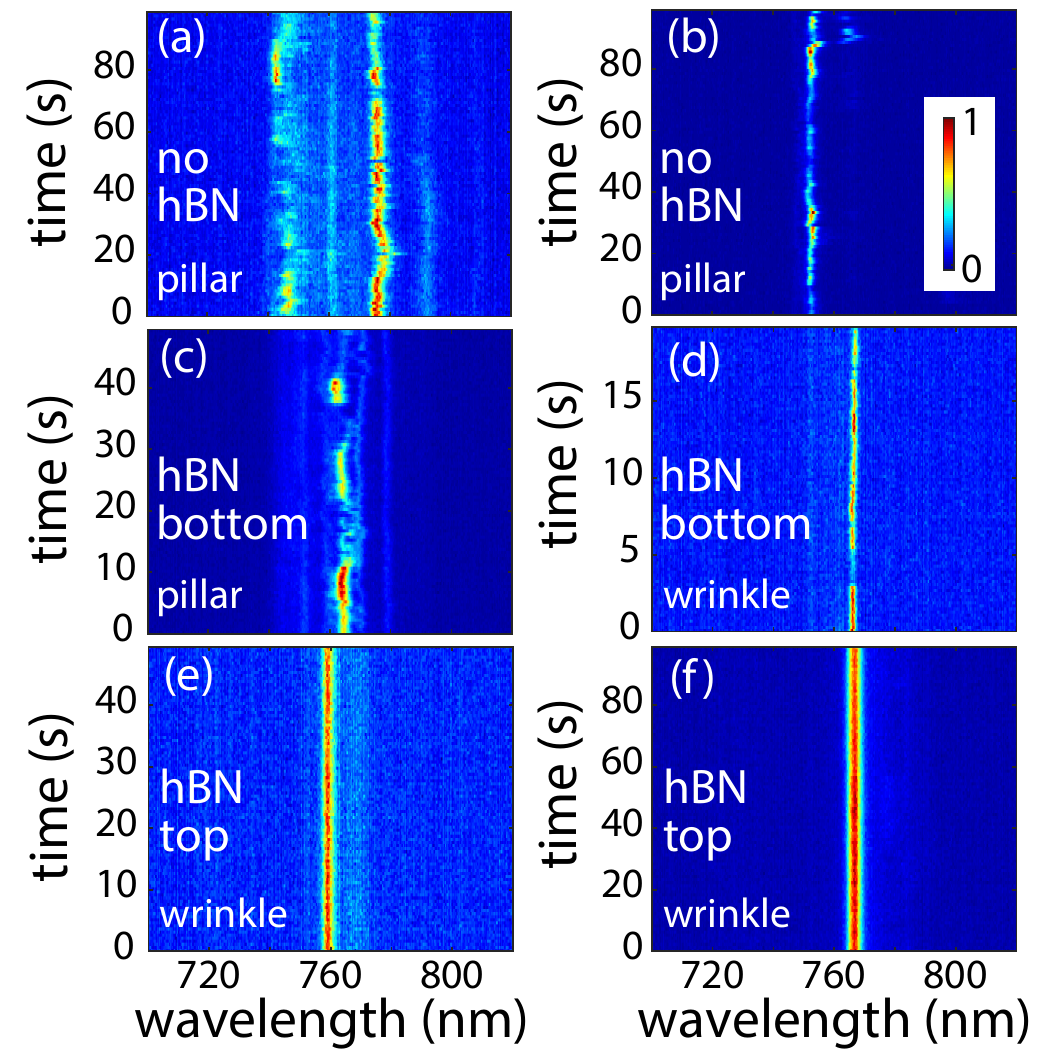}
  \caption{Time traces of \wse quantum emitters to show the effect of \wse surrounding on emitter stability. a-b) Spectra from two emitters directly on SiO$_2$ nanopillars (sample~2) showing substantial amount of spectral instability.  c-d) Spectra of two emitters with a thin hBN layer between SiO$_2$ and \wse (sample~1) stemming from a pillar and wrinkle, respectively. Data from c) corresponds to the pillar QE shown in Fig.~\ref{fig:fig1}(e) and data from d) corresponds to the wrinkle QE shown in Fig.~\ref{fig:fig1}(d). e-f) Spectra of two wrinkle emitters from sample~2 with a thin hBN layer capping WSe$_2$. These emitters are labeled F and G in Fig.~\ref{fig:fig2}(e), respectively. The latter case makes \wse emitters both spectrally and temporally stable, while uncapped \wse quantum emitters are prone to more spectral diffusion and blinking.}
  \label{fig:fig3}
\end{figure}

Another aspect that influences the applicability of single-photon sources is the emitter's spectral and temporal stability. Spectral diffusion and blinking is commonly observed in all solid-state emitters because of fluctuations in the electrostatic environment.  Such spectral instabilities are detrimental because they reduce the two-photon indistinguishability and overall excitation/collection efficiency with resonant excitation or spectral filtering \cite{Lodahl2015interfacing}. Spectral wandering is commonly reported for \wse QEs on SiO$_2$ substrates, however, blinking is seldom observed \cite{srivastava2015optically, chakraborty2015voltage}. Hexagonal boron nitride has been previously used with \wse as a buffer layer to isolate from the substrate \cite{kumar2016resonant}, however the effect of the substrate on QE stability has not been fully understood. Tonndorf \emph{et al.} \cite{tonndorf2015single} studied the effect of using hBN versus SiO$_2$ substrates and found no clear influence on QE stability although hBN was found to increase non-radiative recombination channels. Substrate engineering with InGaP was also shown to yield \wse QEs with no detectable spectral diffusion \cite{iff2017substrate}. For strain-activated QEs, pillars with a 2:1 aspect ratio (200~nm tall, 100~nm width) create less emitters per pillar with low spectral wandering and blinking whereas 1:1 aspect ratio pillars host more QEs, which tend to display substantial blinking and spectral diffusion over time \cite{palacios2017large}.

In the process of working with hBN/\wse stacks on SiO$_2$ substrates, we have also examined the effect of using an hBN passivation layer (hBN) on the spectral stability of QEs. Figure~\ref{fig:fig3} shows a spectral time trace for three different configurations of \wse QE environments, in particular with respect to the presence of hBN. Figure~\ref{fig:fig3}(a-b) shows two QE spectra from sample~2 taken from two nanopillars in an area of the sample where hBN does not cover the \wse monolayer. Overall, we observe that all emitters in direct contact to SiO$_2$, without hBN either above or below, show significant spectral wandering and blinking. Figure~\ref{fig:fig3}(c-d) shows two QE spectra from sample~1 where the \wse is not capped, but hBN separates the QEs from the SiO$_2$ substrate. Data from \ref{fig:fig3}(c) and (d) comes from a pillar QE and wrinkle QE and correspond to the same QEs shown in Fig.\ref{fig:fig1}(e) and (d,f), respectively. Finally, Fig.~\ref{fig:fig3}(e-f) plots data for two wrinkle QEs from sample~2 (spots F an G) that are capped by a thin hBN layer. 

Between the two latter configurations, we find that using hBN as a capping layer instead of a separation layer from the substrate creates the most spectrally stable emitters. On sample~2,  wrinkle-based emitters show no sign of blinking or spectral wandering, and we notice that about 75~\% of pillar QEs are spectrally stable. One difference between the two cases is the fact that wrinkles elevate \wse from the substrate, meaning that wrinkle-based QEs do not touch the substrate whereas nanopillars emitters likely do. Direct contact of \wse with the substrate in Fig.~\ref{fig:fig3}(a-b) is clearly detrimental to the emitter stability and could explain why pillar emitters on average have worse spectral stability than wrinkle-based emitters. Another strong indication that wrinkle emitters are not spectrally wandering is the fact that a Lorentzian function fits very well the QE line in multiple cases (as observed when calculating $r_0$ for sample~2). Spectral wandering on a time scale shorter than the integration time would distort the Lorentzian lineshape and make the fit impractical. Moreover, the study of spectral wandering should be refined in a further study but is currently limited to the larger linewidth of the emitters at 10~K (FWHM~$>1$~meV).

\section*{Three-dimensional confinement of excitons from hBN wrinkles}

\begin{figure*}[t!]
  \centering
  \includegraphics[width=\linewidth]{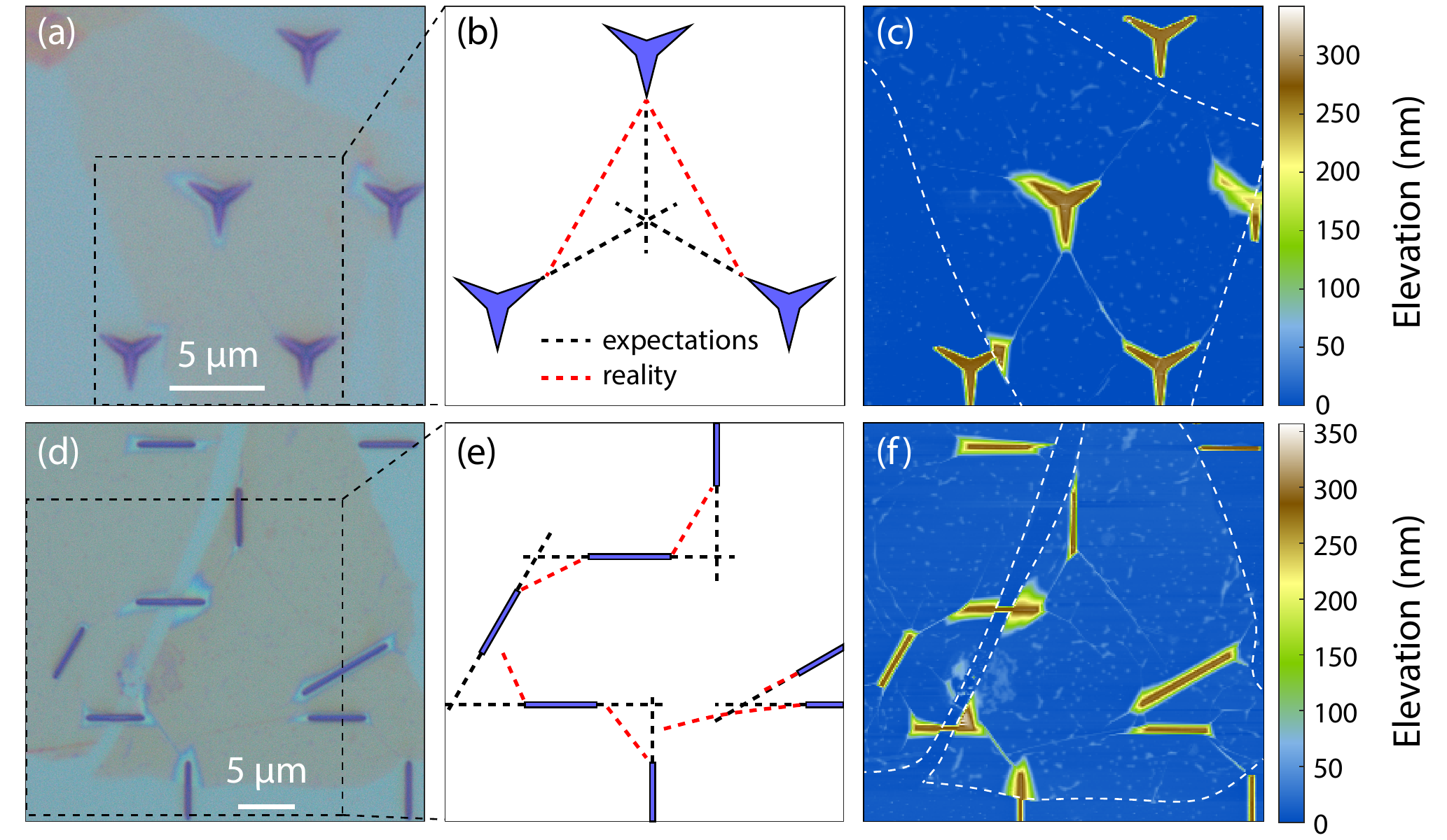}
  \caption{Device design towards wrinkle control. a-b-c) Design featuring narrow/sharp ridges to guide the propagation of wrinkles. d-e-f) Expansion of the previous design with varying angle between propagating ridges. c) and f) show the AFM images of the samples where the hBN wrinkles are clearly visible. The hBN flakes contours are indicated in dashed lines. b) and e) show the expected wrinkles (dashed black lines) and the observed ones (dashed red lines) as measured with AFM. For both samples, observations are similar: wrinkles bind between nearest neighbors and do not cross.}
  \label{fig:fig4}
\end{figure*}

Multiple experiments suggest that single-photon emitters in \wse are quantum dot like, i.e., originating from the quantum confinement of single excitons \cite{shepard2017nanobubble, darlington2020imaging, branny2017deterministic, palacios2017large}. Tensile strain locally shrinks the \wse bandgap energy, which may lead to a potential well that hosts discrete energy levels similarly to semiconductor quantum dots \cite{Santori2001}. 
Some theoretical works speculate that tensile strain gradient serves to funnel excitons towards a defect, the presence of which being necessary for single-photon emission \cite{zheng2019point, linhart2019localized, dang2020identifying}. Despite the difference in the underlying capturing process of the exciton, controlling the size scale over which strain is applied to \wse is equally crucial in both cases. When using nanopillars with diameter $>100$~nm to purposefully strain WSe$_2$, it is likely that confinement over the full size scale of the nanopillar is likely not happening. Based on the nanopillar data of Fig.~\ref{fig:fig1}(c), Fig.~\ref{fig:fig2}(e) and other works \cite{palacios2017large, branny2017deterministic}, we hypothesize that multiple confining sites are created around the nanopillars due to a complex and uncontrolled strain profile much larger than the size scale suitable to create quantum confinement \cite{Lodahl2015interfacing}. As a result, nanopillars larger than few tens of nanometers are likely to host several QEs within the excitation/emission spot.

\begin{figure*}[t!]
  \centering
  \includegraphics[width=\linewidth]{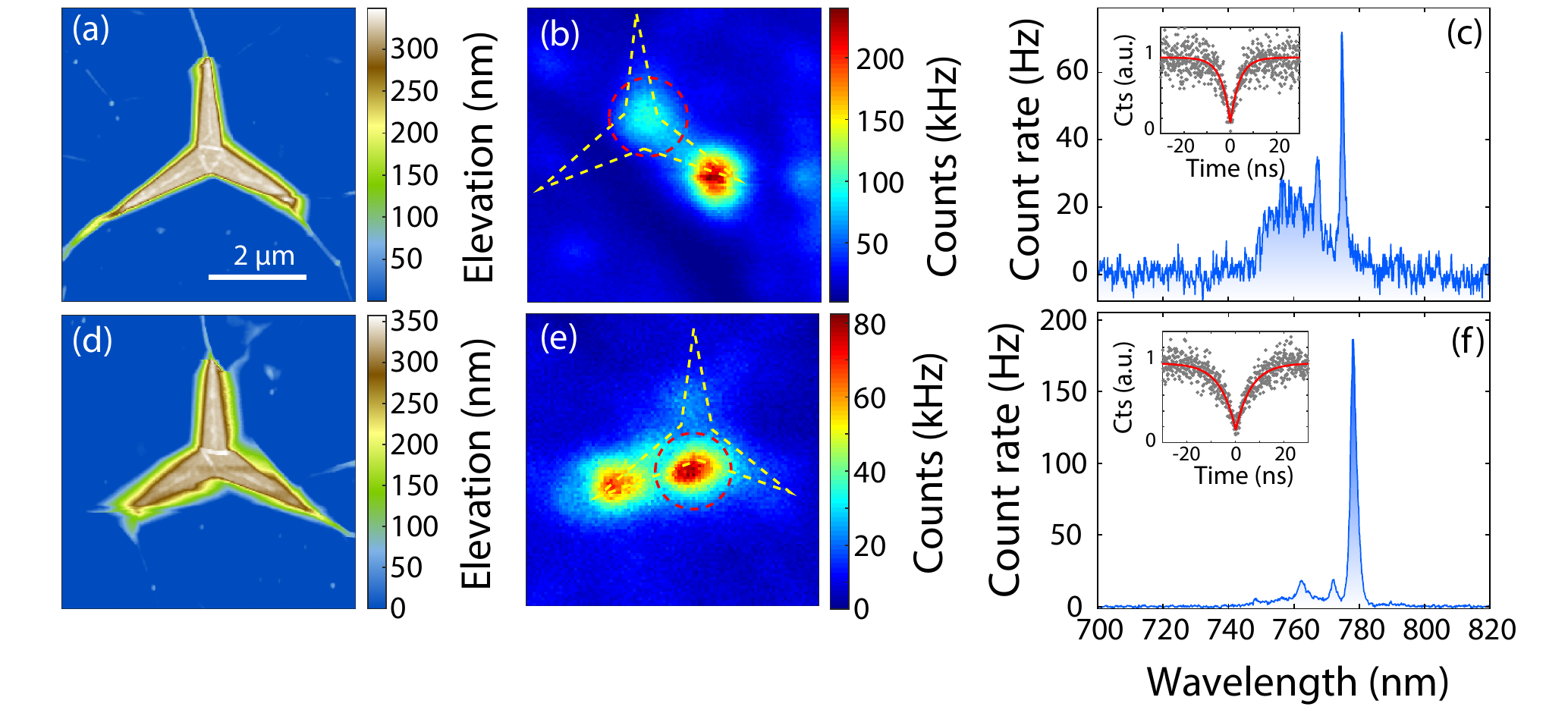}
  \caption{\wse quantum emitters characterization on star-triangle nanostructure. a) Atomic force microscopy image of a WSe$_2$/hBN sample. Wrinkles form away from each arm as well as on top of the structure in a triangular shape. b) Photoluminescence map of the area shown in a) with a 637~nm laser at 2 \micro W. c) Spectrum and filtered $g^{(2)}(\tau)$ of the luminescent center (circled in red). d-e-f) Similar data for a different star-triangle structure. The spectrum corresponds to the bright center (circled in red).}
  \label{fig:fig5}
\end{figure*}

Figures~\ref{fig:fig1}-\ref{fig:fig2} have shown that the use of a thin hBN flake over a nano-patterned substrate to regulate the amount of strain in \wse may be a better choice to create single \wse QEs than using substrate features directly to strain WSe$_2$. However, the positioning of QEs along hBN wrinkles in these samples is random, whether at a kink or in nano-bubbles. The remainder of this paper focuses on ways to create hBN wrinkles so as to control the positioning of QEs. A single wrinkle only provides one in-plane dimension of confinement, similar to semiconductor nanowires. We have considered several ways to extend the confinement provided by wrinkles to two in-plane dimensions.  As a first approach, we designed the structures shown in Fig.~\ref{fig:fig4}, which are meant to have two propagating wrinkles cross and create a high strain point. Figure~\ref{fig:fig4}(a,b,c) show an optical micrograph and an AFM image of sample~3a containing star-triangle structures overlaid by a thin hBN flake. In most cases, wrinkles connect nearest-neighbor ridge structures instead of propagating independently and in the direction of the ridge, i.e., creating a crossing of two wrinkles is not straightforward. Moreover, since the transfer process is manually operated and uses a visco-elastic PDMS stamp, the details of the transfer matter: propagation direction, transfer speed and amount of pressure applied all influence the outcome. In Fig.~\ref{fig:fig4}(d,e,f), we show sample~3b containing propagating ridges with varying separation angles. At shallow angles (30 degrees), the ridges do form wrinkles that propagate in the direction of the ridge, however, rather than crossing, the two wrinkles avoid one another, ending up propagating in a roughly parallel path. At larger angles, the wrinkles bind directly between nearest-neighbor ridge ends as seen in Fig.~\ref{fig:fig4}(c). Figures.~\ref{fig:fig4}(b) and (e) sum up our expected and observed wrinkle propagation direction as black and red dashed lines, respectively. We conclude that creating two independent hBN wrinkles to achieve a wrinkle crossing is not feasible. Other approaches may be better, e.g., a hybrid sample designed such that a single hBN wrinkle crosses a lithographically defined shallow feature to bring additional strain on a single wrinkle.

Several groups have demonstrated that strain activation of QEs extends beyond the use of nanopillars, by using e.g., slotted waveguides \cite{kern2016nanoscale, blauth2018coupling} or hBN nano-bubbles \cite{shepard2017nanobubble}. We also exemplify the universality of strain induction of QEs in \wse monolayers by using a star-triangle patterned substrate (sample~3a) with a WSe$_2$/hBN heterostructure. The heterostructure is transferred on top of 12 star-triangle shapes: we record bright distinguishable emission centers on every structure with an average of 2.9 bright centers per star-triangle feature and an average of 3.1 emitters per bright center (out of 21 studied). Figure~\ref{fig:fig5} shows results from two star-triangles. The AFM images in Fig.~\ref{fig:fig5}(a)(d) show how the heterostructure conforms to the star-triangle shape. Interestingly, we notice a wrinkle pattern on top of each star that resembles a triangle. In both instances, bright emission centers appear at the tip of one branch and from the center of the structure, see PL map in Fig.~\ref{fig:fig5}(b)(e). The star-triangle shape is indicated in a yellow dashed line and was inferred from cross referencing the PL-map, AFM scan and optical image of the sample. The associated spectra and filtered $g^{(2)}(\tau)$ are shown in Fig.~\ref{fig:fig5}(c)(f). The spectra are reminiscent of those collected from pillars in samples~1 and 2, where the background level is rather pronounced. The bright emission centers recorded at the center of the star-triangle may be coming from the wrinkles that form on top of the stars in the AFM images. However, we speculate that the tightness of the wrinkles along with the complexity on top of the star structures creates a complex strain profile, hence the more crowded spectra. The filtered $g^{(2)}(0)$ extracted from the fit are 0.14 and 0.16. This sample shows that working with any structure that produces local strain is likely to create QEs in \wse and that the spectral quality of emitters can vary based on transfer process, hBN thickness and morphology.

\section*{Conclusion}
We have reported the observation of \wse quantum emitters that form from strain created along hexagonal boron nitride wrinkles. These emitters have been shown to be frequently isolated from other emitters and background emission. Quantum emitters found along hBN wrinkles typically have a  peak-to-background ratio exceeding 90~\%, which defines the fraction of a QE intensity over all the collected light in a 3~nm spectral window. Remarkably, an analysis of the unfiltered spectrum shows that on average, bright emission centers associated with QEs along hBN wrinkles contain four times as little residual emission (background and other emitters) compared to emitters formed directly on substrates features such as a nanopillar. We also found that using a capping hBN layer to protect   \wse QEs from the environment confers the best configuration to suppress spectral wandering and blinking in our study. In particular, we attribute the ideal spectral stability of wrinkle QEs to the fact that hBN wrinkles elevate \wse from the SiO$_2$ substrate, while pillar emitters remain in contact with the substrate. Finally, we have studied alternative substrates to control hBN wrinkle propagation and conclude that crossing two wrinkles to create a high-strain point is not feasible. We suggest for a future work that propagating a hBN wrinkle over a perpendicularly defined shallow structure could create a high-strain point along the wrinkle, thereby creating a \wse QE with position control. Our combined results -- single QE created per site, low surrounding background, and spectral stability -- make wrinkle-based \wse QEs promising for applications requiring highly pure single-photon sources \cite{Lodahl2015interfacing, obrien2009photonic, gisin2002quantum, kimble2008quantum}. Further research that enables the use of the valley properties of \wse excitons would bring even more functionality to the source, such as control over photon polarization \cite{mak2016photonics}.

\section*{Author's contributions}
RSD, TAV and GDF designed the experiment and wrote the manuscript. RSD and TAV prepared samples, collected and analyzed data. AM and ANV performed additional measurements on the samples. ZW, KFM and JS helped with sample preparation.  All authors reviewed the manuscript.

\section*{Data Availability}
The data that support the findings of this study are available from the corresponding author upon reasonable request.

\section*{acknowledgments}
This work was primarily supported by an AFOSR MURI (FA9550-18-1-0480) with partial support from the Cornell Center for Materials Research (CCMR) with funding from the NSF MRSEC program (DMR-1719875), NSF Career (DMR-1553788) and AFOSR FA9550-19-1-0074. We also acknowledge facility use from the CCMR and from the Cornell NanoScale Facility, a member of the National Nanotechnology Coordinated Infrastructure (NNCI), which is supported by the National Science Foundation (NNCI-1542081).

\section*{Appendixes}
\appendix

\section{Peak-to-background ratio calculation details}
\label{sec:SI:r0}

\begin{figure}[h!]
  \centering
  \includegraphics[width=0.5\linewidth]{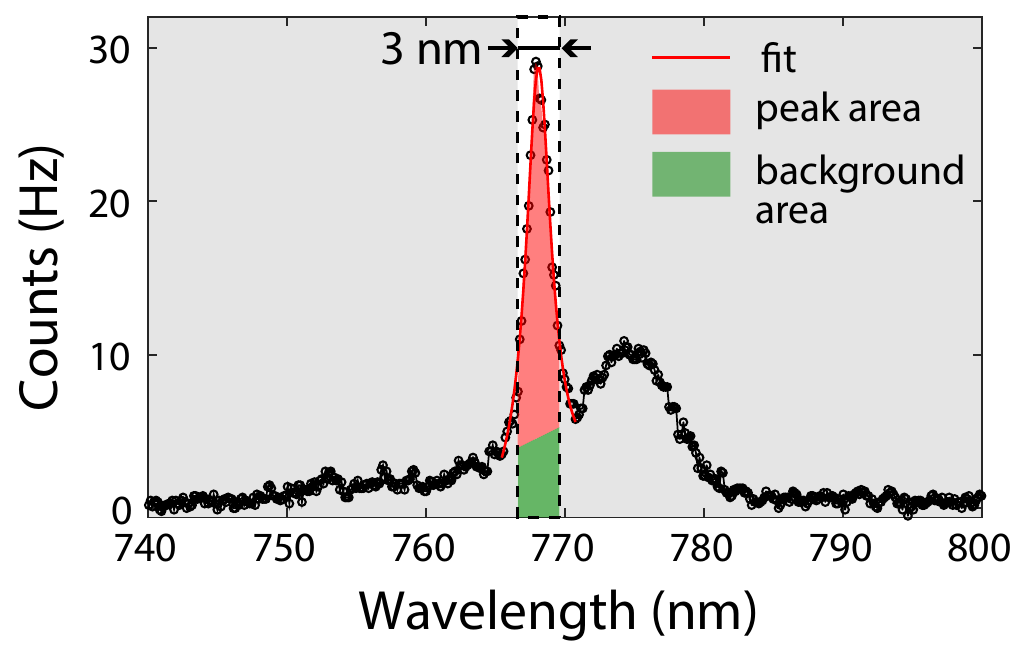}
  \caption{Spectrum fitted with a Lorentzian lineshape to exemplify how the peak-to-background ratio $r_0$ is estimated. The fit is used primarily to extract the wavelength-dependent background level within the peak. The dashed box corresponds to the pixels of interest to calculate $r_0$. Here $r_0 = 0.76\pm0.02$.}
  \label{fig:figSI1}
\end{figure}

Figure~\ref{fig:figSI1} shows the spectrum of wrinkle QE B from sample~2 (see Fig.~\ref{fig:fig2}(e)) along with a fit of the  QE line. The fitted function is composed of a Lorentzian lineshape $I_\text{peak}(\lambda)$ plus a wavelength-dependent background $I_\text{background}(\lambda)$ as discussed in the main text. The peak area and background are calculated by summing the pixels of the red and green area within a spectral window of 3~nm indicated as a dashed box. In this example, the computed $r_0 = 0.76\pm0.02$.


%

\end{document}